\documentclass[acmsmall,screen,authorversion,nonacm]{acmart}

\usepackage{amsmath,amsfonts}
\usepackage{graphicx}
\usepackage{subcaption}
\usepackage{xspace}
\usepackage{booktabs}
\usepackage{array}
\usepackage[shortlabels]{enumitem}
\usepackage{url}
\usepackage{algorithmic}
\usepackage{microtype}
\usepackage{balance}
\usepackage{tikz}

\setcitestyle{nocompress,noadjust}
\usepackage[noabbrev,nameinlink,capitalize]{cleveref}

\newcommand{\df}[1]{{\it #1}}
\newcommand{\bin}[1]{{\texttt{#1}}\xspace}
\newcommand{\alg}[1]{{\textsl{\textsf{#1}}}\xspace}
\newcommand{\unknown}{{\ensuremath{\varnothing}}}

\DeclareMathOperator*{\tsp}{TSP}
\DeclareMathOperator*{\cnt}{cnt\xspace}

\begin{document}

\title{Stale Profile Matching}

\author{Amir Ayupov}
\orcid{0009-0009-3211-2000}
\affiliation{%
	\institution{Meta}
	\city{Menlo Park, CA}
	\country{USA}
}
\email{aaupov@meta.com}

\author{Maksim Panchenko}
\orcid{0009-0009-5155-8139}
\affiliation{%
	\institution{Meta}
	\city{Menlo Park, CA}
	\country{USA}
}
\email{maks@meta.com}

\author{Sergey Pupyrev}
\orcid{0000-0003-4089-673X}
\affiliation{%
	\institution{Meta}
	\city{Menlo Park, CA}
	\country{USA}
}
\email{spupyrev@meta.com}

\authorsaddresses{}

\begin{abstract}
Profile-guided optimizations rely on profile data for directing compilers to generate optimized code.
To achieve the maximum performance boost, profile data needs to be collected
on the same version of the binary that is being optimized. In practice however, there is typically
a gap between the profile collection and the release, which makes a portion of the profile invalid
for optimizations. This phenomenon is known as profile staleness, and it is a serious practical problem
for data-center workloads both for compilers and binary optimizers.

In this paper we thoroughly study the staleness problem and propose the first practical solution
for utilizing profiles collected on binaries built from several revisions behind the release.
Our algorithm is developed and implemented in a mainstream open-source post-link optimizer, BOLT.
An extensive evaluation on a variety of standalone benchmarks and production services indicates
that the new method recovers up to $0.8$ of the maximum BOLT benefit, even when most of the input
profile data is stale and would have been discarded by the optimizer otherwise.
\end{abstract}

\maketitle

\section{Introduction}
Mobile applications and ubiquitous AI workloads became an essential part of everyday life,
making it crucial to optimize for their efficiency and reliability. Profile-guided optimizations (PGO),
also called feedback-driven optimizations,
are a collection of compiler techniques that use runtime information collected via profiling for improving
the program execution. Modern PGO is successful in speeding up server workloads by providing up to a
double-digit percentage boost in performance~\cite{PANO19,Prop21,HMPWY22}. Similarly, PGO applied for mobile applications
reduces their size and the launch time, which directly impacts user experience, and hence, user retention~\cite{LFZLS22,LHT22,HLMP23}.
Therefore, PGO is nowadays a standard feature in most commercial and open-source compilers.

Traditionally, PGO is a combination of compiler optimizations, including function inlining, register allocation, and
code layout. It relies on execution profiles of a program, such as the execution frequencies of basic
blocks and function invocations, to guide compilers to optimize critical parts of the program
more effectively. While many optimization passes can be applied without
profile data, knowing the execution behavior of a program allows
the compiler to generate a significantly optimized code. Early efforts on PGO were implemented via
compile-time instrumentation, which injects code
to count the execution frequencies of basic blocks, jumps, and function calls. This approach, however,
not only complicates the build process, but also incurs significant performance overhead, which may alter the program's default
behavior and make the collected profiles non-representative. Later works on PGO employ
sampling-based approach that rely on hardware performance counters available in modern CPUs,
such as Intel's Last Branch Records (LBR). Sampling-based PGO enables profiling in the production environment
with a negligible runtime overhead, although the collected profiles may require an extra
post-processing adjustment~\cite{Lee15,Zhou2016,HMPWY22}.
AutoFDO~\cite{CML16}, BOLT~\cite{PANO19}, and Propeller~\cite{Prop21} are examples of frameworks
for optimizing data-center workloads based on the technology.

\paragraph{Continuous Profiling}
While PGO systems have been successfully deployed at scale in production, there is one challenge
that often remains overlooked. In order to achieve the maximum performance boost, profile data supplied to
PGO needs to be statistically representative of the typical usage scenarios. Otherwise, an optimization
has the potential to regress the performance instead of improving it. Consider for example, a function
that is hot according to a profile but is rarely executed in real-world usage. A compiler may decide to
inline the function, which might cause a worsened instruction cache performance due to the increased code size
while not providing any runtime benefits~\cite{Chen93,TGS22}.

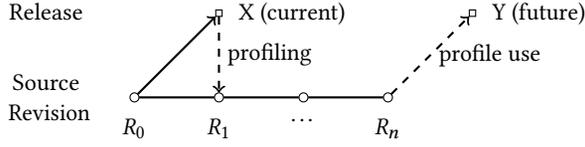
\begin{figure}[!tb]
\usetikzlibrary {positioning,shapes.misc}
\small
\begin{tikzpicture}[node distance=5mm,
                    src/.style={
                      circle,
                      scale=0.4,
                      draw
                    },
                    rel/.style={
                      scale=0.4,
                      draw
                    }
                    ]
  \node (srcrev) [text width=1cm,align=center] {Source\\Revision};
  \node (rel) [text width=1cm,align=center,above=0.5cm of srcrev] {Release};
  \node (crc1) [src,right=of srcrev] {};
  \node (a) [below=3pt of crc1] {$R_0$};
  \node (crc2) [src,right=1cm of crc1] {};
  \node (b) [text width=1cm,align=center,below=3pt of crc2] {$R_1$};
  \node (crc3) [src,right=1cm of crc2] {};
  \node (c) [text width=1cm,align=center,below=3pt of crc3] {\dots};
  \node (crc4) [src,right=1cm of crc3] {};
  \node (d) [text width=1cm,align=center,below=3pt of crc4] {$R_n$};
  \node (crc5) [src,draw=none,right=1cm of crc4] {};
  \node (bx1) [rel,above=1cm of crc2] {};
  \node (x) [right=3pt of bx1] {X (current)};
  \node (bx2) [rel,above=1cm of crc5] {};
  \node (y) [right=3pt of bx2] {Y (future)};
  \path[thick,->] (crc1) edge (bx1);
  \path[thick,-]
  (crc1) edge (crc2)
  (crc2) edge (crc3)
  (crc3) edge (crc4)
  ;
  \path[thick,dashed,->]
  (bx1) edge node [right] {profiling} (crc2)
  (crc4) edge node [right] {profile use} (bx2)
  ;
\end{tikzpicture}
\caption{Continuous profiling causes a mismatch between revisions used to produce the
  profile ($R_0$) and to which the profile is applied ($R_n$).}
\label{fig:cicd_gap}
\end{figure}

In order to collect a representative profile, continuous profiling systems are employed at large software companies~\cite{GWP10}.
These systems collect a sampled profile from the fleet and aggregate them for subsequent use in compiler optimizations.
However, in such systems, the profile is always lagging the most recent source~\cite{CML16}; see \cref{fig:cicd_gap}.

Another related issue occurs when the source code of a program is modified right before the release (which is known as a \df{hotfix}).
Even when such changes touch a single line of the source code, they can lead to substantial changes in the generated machine
code, making previously collected profiles \df{stale}, that is, invalid for optimizations.
For these reasons, modern PGO systems assume that profiles are collected
on exactly the same version of a program that is being optimized, whereas stale profile data is completely discarded.

How severe is the staleness problem in practice? One would assume that the code for large programs, referred to as \df{applications}
or simply \df{binaries},
does not change too frequently and the majority of the profile data remains unchanged between consecutive releases.
However, empirical data collected by the BOLT binary optimizer used for optimizing large-scale services~\cite{PANO19,PASO21} 
disagrees with the intuition. We record $70\%$ stale samples between two (weekly) releases
for one large-scale service, and over $92\%$ staleness after only a three-week delay of updating the profiles for another one.
Thus, the benefit of applying BOLT reduces by two thirds for the former service and almost diminishes for the latter one.
The phenomena are explained as follows. Profiling information is captured at the machine code level and stored at the function
granularity. If a function is unchanged between two releases, the profile data can be reused for optimizations. In contrast,
when the content of the function is modified, then the profile data becomes stale. Generally even innocent changes,
such as adding or removing padding, result in modified jump and call addresses and their offsets from the beginning
of the function, which leads to invalid profiles. In addition, function inlining results
in waterfall modifications at all call sites.

With these problems in mind, we develop a novel approach for \df{stale profile matching}, which is a technique for
adjusting and re-using profile information gathered on a pre-release version of a program. In a sense, we relax
the requirement of profiling and releasing the same version of a binary. 
That significantly simplifies the development process and
eases the adoption of PGO systems in the real-world environment. As our key example, we show that for the large-scale
\bin{clang} binary, we recover $0.78$ of PGO benefits even when over $90\%$ of its profile data (collected on
a six-month-old release) is stale. This is equivalent to a $5.9\%$ absolute speedup of the binary on top
of the state-of-the-art optimizations.

\paragraph{Our Contributions.}
The primary contributions of the work are summarized as follows.

\begin{itemize}[leftmargin=*]
	\item Firstly, we thoroughly investigate profile staleness in PGO and propose a formal
	model for the problem, capturing practical constraints and objectives.
	Then, we develop a novel two-stage algorithm for the problem and demonstrate how it is applied
	for reducing profile staleness.

	\item Secondly, we present an implementation of the new approach in a post-link binary optimizer, BOLT,
	which is a part of LLVM~\cite{LA04}.
	The implementation is fully integrated into the main branch of the optimizer and requires no changes
	to the workflow for BOLT users. The algorithm is relatively simple and efficient, being able to process
	large production binaries without noticeable runtime overhead.

	\item Finally, we extensively evaluate the new approach on a variety of benchmarks, including
	five large open-source binaries and four production workloads at Meta. The experiments indicate that
	the new method recovers $0.6{-}0.8$ of the maximum BOLT speedup, even when most of the input
	profile data is stale.
\end{itemize}

\begin{figure*}[!tb]
	\centering
	\begin{subfigure}[t]{0.48\columnwidth}
		\includegraphics[width=\columnwidth]{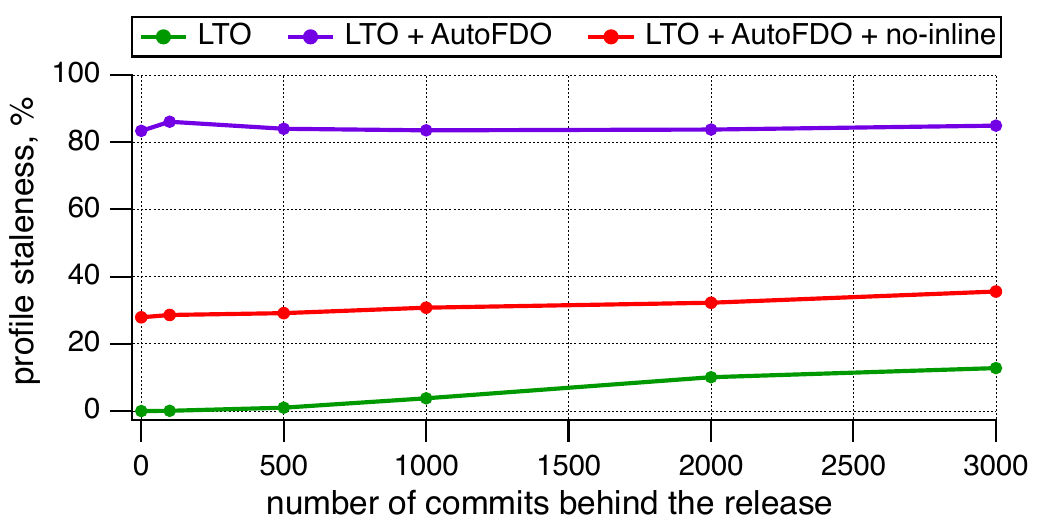}
		\caption{The percentage of stale (invalid) samples in the profile collected
			on a binary several commits behind the release}
		\label{fig:clang_staleness}
	\end{subfigure}
	\hfill
	\begin{subfigure}[t]{0.48\columnwidth}
		\includegraphics[width=\columnwidth]{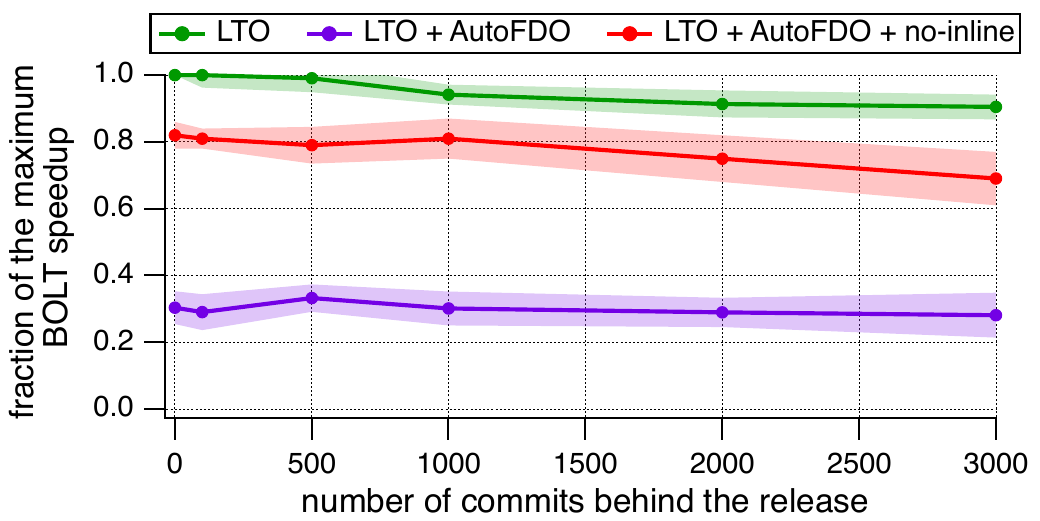}
		\caption{Performance regression of \bin{clang} optimized with BOLT utilizing stale profile data}
		\label{fig:clang_perf}
	\end{subfigure}
	\caption{Investigation of profile staleness for the \bin{clang} binary (\texttt{release\_15}) built
		in different modes.}
\end{figure*}

We emphasize that existing literature on the topic is rather sparse with only two works tackling the problem.
Wang, Pierce, and McFarling~\cite{WPM00} design a binary matching tool for stale profile propagation
for Microsoft NT (2000) applications. A very recent preprint of Moreira, Pereira, and Ottoni~\cite{PMO23}
independently explore a method for using a stale profile in binary optimizers.
Both methods rely on a hash-based matching between basic blocks of a program. As we argue in \cref{sect:invest}
and experimentally demonstrate in \cref{sect:eval}, such approaches are unable to provide adequate
performance benefits for real-world instances. In contrast, our work presents the first two-stage
algorithm comprised of \df{matching} and \df{inference} (defined in \cref{sect:algo}), which produces
optimal and near-optimal results in practice.

We also stress that while our implementation and evaluation is done for a mainstream post-link binary
optimizer, BOLT, the approach discussed in the paper is general enough to be used in other contexts.
In particular, we believe that it is possible to adapt the technique to Propeller~\cite{Prop21}
or use it in conjunction with LLVM's profile-guided optimizations~\cite{CML16}.

\paragraph{Paper Organization.}
The rest of the paper is organized as follows.
We first investigate the phenomena of profile staleness in \cref{sect:invest} and
analyze the limitations of existing PGO tools. Building upon the knowledge, we develop
a novel approach for re-using stale profiles. \cref{sect:algo} describes the two
components of our methodology, \df{matching} and \df{inference}, and describes an
implementation in an open-source binary optimizer, BOLT, developed on top of LLVM.
Next in \cref{sect:eval}, we provide a detailed experimental evaluation of the
algorithm on a rich collection of open-source binaries and production workloads.
\cref{sect:related} discuss related work in the area.
We conclude the paper and propose possible future directions in \cref{sect:conclusion}.

\section{Investigating Profile Staleness}
\label{sect:invest}

To shed light on the staleness problem, we investigate differences between consecutive releases of a binary optimized with BOLT. We chose to experiment with a standalone binary of the \bin{clang} compiler, which has a relatively large code size, and can be easily integrated with various PGO technologies. As a baseline, we utilize \texttt{release\_15} of the binary cut at January 2023, and consider up to $3000$ commits prior to the release so that the most stale data ($3000$ commits behind \texttt{release\_15}) corresponds to June 2022.

Our first experiment, whose results are visualized in \cref{fig:clang_staleness}, measures the percentage of stale samples
in the profile data collected on a binary built for the source code corresponding to $x$ commits behind \texttt{release\_15}. We vary
$x$ between $0$ and $3000$ so that $x=0$ corresponds to the release. The level of staleness depends on how the baseline binary is built.
In the ``simplest'' mode with the optimization level \texttt{O3} and link-time optimizations (\texttt{LTO}) enabled, profile staleness starts at $0\%$ for $x=0$, that is, the profile has no stale samples. The quality of the profiles slowly degrades, as the gap between the release and the profiled binary increases. In this setup, collecting profiles on a 3-month-old binary ($500$ commits behind the release) invalidates less than $3\%$ of samples, while profiling a 6-month-old binary ($x=3000$) yields $12.8\%$ stale samples in the profile.
The results look quite different when we start utilizing compiler's PGO for building the baselines binary; in the evaluation we
experiment with sampling-based AutoFDO~\cite{CML16}. Profile staleness reported by BOLT reaches $83\%$ even when we rebuild \bin{clang}
using the same source code (the rebuild process, however, involves re-running the AutoFDO step);
the value remains stable across the experiment. Such staleness almost entirely invalidates the profile data supplied to BOLT.
To further understand the issue, we repeated the experiment by building the binary with inlining disabled, that is, using the  \texttt{O3+LTO+AutoFDO+no-inline} mode. In that experiment, profile staleness is recorded at the initial value of 28\% and slowly grows
with the number of commits, $x$.

\cref{fig:clang_perf} illustrates the performance impact of using the stale profile data. The plot reports the fraction of the maximum
speedup on the \bin{clang} binary achieved by BOLT utilizing stale profiles, relative to the maximum speedup it can achieve using
\df{fresh} profile data collected on the same revision.
In the evaluation, the BOLT speedup is recorded at $28\%$ on top of the non-BOLTed counterpart in the
\texttt{O3+LTO} mode and $12\%$ in the \texttt{O3+LTO+AutoFDO} mode; the values agree with the original speedups reported
by the BOLT team in \cite{PANO19,PASO21}. As expected, the impact of applying BOLT is inverse proportional to profile
staleness. For the simpler \texttt{O3+LTO} build mode, stale profile data yields a modest $2\%{-}3\%$ regression. However,
for the practical \texttt{O3+LTO+AutoFDO} mode, utilizing a 6-month-old profile results in a substantial regression:
instead of the maximum possible speedup of $13\%$, BOLT realizes only $3.5\%$, which is an equivalent of value $0.26$ on the plot.

\begin{figure}[!tb]
	\centering
	\includegraphics[width=0.5\columnwidth]{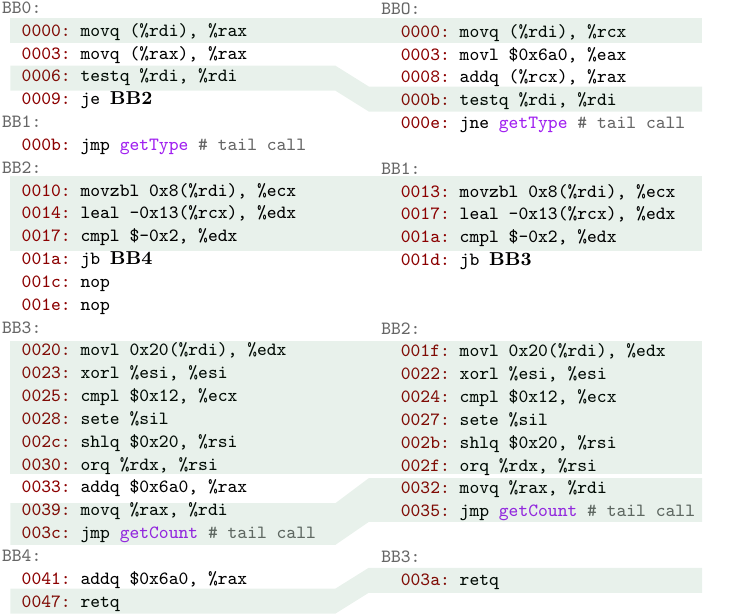}
	\caption{A function in \bin{clang} built with AutoFDO in
		the profiled (left) and the release (right) binaries.}
	\label{fig:stale_ex}
\end{figure}

In order to further understand the problem, we analyze individual functions in \bin{clang}, whose
profile data is marked stale. We identified three primary reasons for profile staleness. Firstly,
developers modify the source code, which directly causes changes in the generated binary. Such changes
include, for example, code being added or removed and function renaming or type changes.
Secondly, we noticed many minor differences in the generated code, such as extra
\texttt{nop} instructions or a different treatment of tail calls. \cref{fig:stale_ex} illustrates one such example with
the code of the same function in the profiled binary (left) and the release one (right).
While it is easy for a human to map the basic blocks between the two versions, BOLT's conservative
strategy is to discard the function profile whenever there is a mismatch in the number of basic blocks or jumps.
Such differences are a result of the existing compiler being non-stable and producing non-identical results
when there are small changes in the source code or in the profile data.
Finally, the above two types of modifications are often amplified by function inlining, which
propagates changes in individual functions across many instances.

\section{A New Approach}
\label{sect:algo}

We assume that a \df{binary} (some representation of a compiled program) is a collection of \df{functions}. Each function
is represented by a directed (possibly cyclic) control-flow graph, denoted $G=(V, E)$.
The vertices of $G$ correspond to basic blocks and
directed edges represent jumps between the blocks. We assume that the graph
contains a unique source, $s^{*} \in V$ (that is, the entry block of
the function) but may have multiple sinks (exit blocks), denoted $T^{*} \subset V$, that are reachable from $s^{*}$ via a directed path.
The binary is associated with a \df{profile} dataset collected on a representative benchmark suite.
The profile contains vertex and edge \df{counts}, $\cnt(v) \ge 0$ for $v \in V$ and $\cnt(u, v) \ge 0$ for $(u, v) \in E$,
that represent the execution counts of basic blocks and jumps, respectively. As noted above, the counts
might be imprecise and serve just as an estimation of actual execution counts during profiling.
The profile may also contain additional meta-data, such as the names of the functions or
the sizes of the basic blocks; refer to \cref{fig:yaml} for an example.

\begin{figure*}[!tb]
	\centering
	\begin{subfigure}[t]{0.48\columnwidth}
		\includegraphics[page=2,width=\columnwidth]{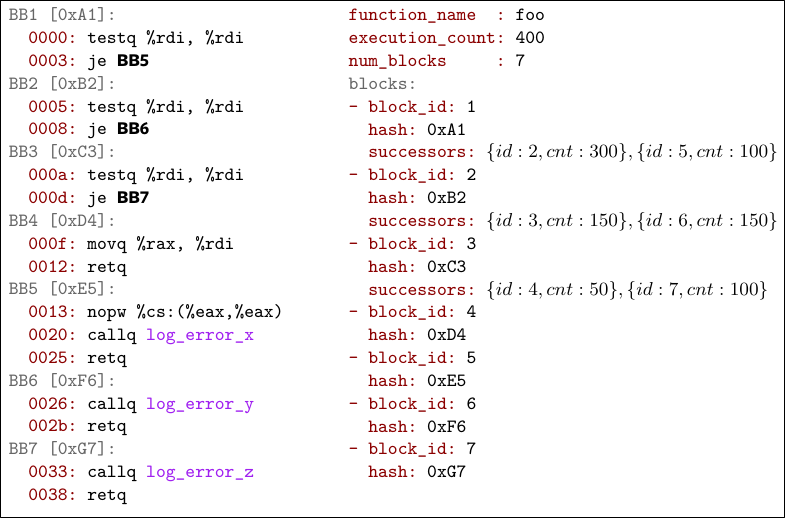}
		\caption{Binary $B_{old}$ (left) with profile $P_{old}$ (right)}
		\label{fig:io1}
	\end{subfigure}
	\hfill
	\begin{subfigure}[t]{0.48\columnwidth}
		\includegraphics[page=3,width=\columnwidth]{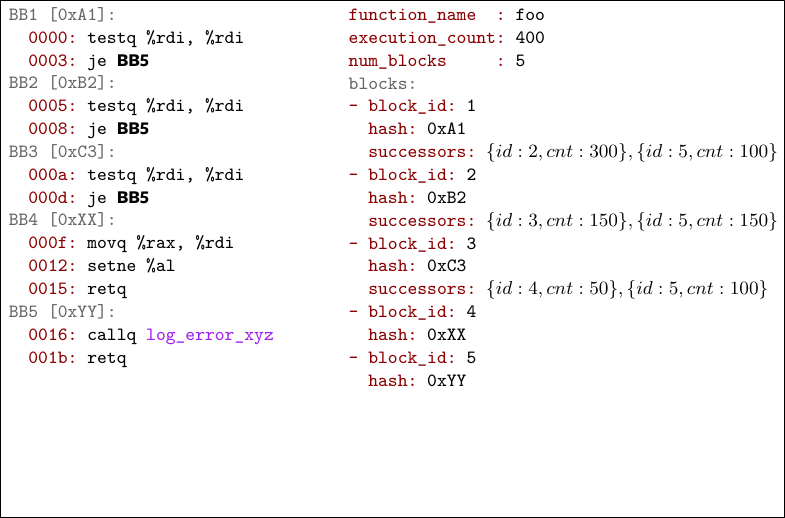}
		\caption{Binary $B_{new}$ (left) with profile $P_{new}$ (right)}
		\label{fig:io2}
	\end{subfigure}
	\caption{A function, \texttt{foo}, modified between two releases (\emph{old} and \emph{new})
		of the binary. $B_{new}$ and $P_{old}$ comprise the input for the stale profile matching problem.
		The goal is to infer a profile, which is as close to $P_{new}$ as possible.}
	\label{fig:io}
\end{figure*}

We assume that there are two releases of a binary:
the older one, $B_{old}$, and the most recent one, $B_{new}$. The binaries are associated with profiles
$P_{old}$ and $P_{new}$, respectively.
We refer to \cref{fig:io} for an example of the same function (\texttt{foo}) in the two releases
of the binary along with the corresponding profiles. In the example we assume
that every basic block is associated uniquely with a \df{hash} value, computed based
on its content, that is, the opcodes and operands of its instructions.
Using the hashes, it is straightforward to find a one-to-one mapping between basic blocks in the
binary and in the profile, as long as the hash values stay the same, that is,
when the content of the basic blocks is unchanged. However, as we argue in \cref{sect:invest}, that is often
not the case. Our goal is to infer the counts of the basic blocks and jumps in binary $B_{new}$ using
profile $P_{old}$. We stress that the strategy prior our work is to discard the profile
and keep unoptimized function \texttt{foo}, since its meta-data (e.g., the number of the basic blocks)
does not match in $B_{new}$ and $P_{old}$.

Now we describe the high-level strategy of our algorithm.
As a pre-processing step independent from stale matching, BOLT finds one-to-one
mapping between the functions in $B_{new}$ and $P_{old}$ based on their names.
It first attempts to match function names exactly. For functions with unique
suffixes such as produced by LLVM LTO for internal linkage symbols
that may drift between compilations, 
BOLT attempts to match using heuristics ignoring the suffix: (i)~match using a function hash,
(ii)~for cases where there is only one function in $B_{new}$ and $P_{old}$ after
stripping the suffix, match them up despite hash mismatch. In the case of the remaining
ambiguity, BOLT ignores the function profile. Hence, 
functions that have been added or deleted are discarded. 

\begin{figure*}[!tb]
	\centering
	\includegraphics[page=1,width=0.75\textwidth]{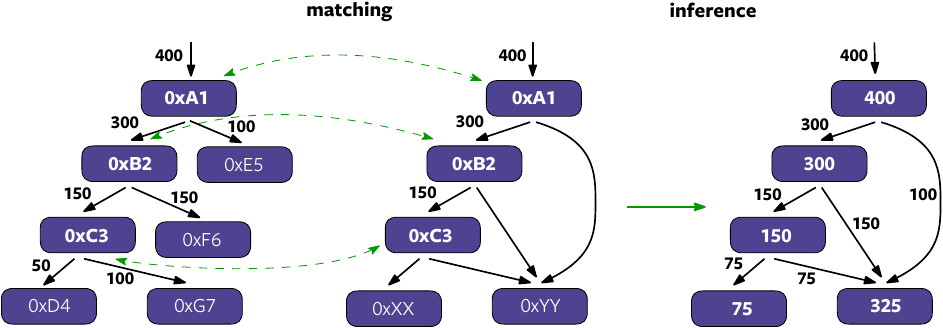}
	\caption{An overview of our algorithm for the stale profile matching problem.}
	\label{fig:algo}
\end{figure*}

The main step of the algorithm
consists of two phases, the \df{matching} phase and the \df{inference} phase.
We refer to \cref{fig:algo} for an overview, which illustrates the process for function \texttt{foo}
from \cref{fig:io}.

The matching phase is used to extract partial information from the profile, $P_{old}$, and assign
\df{initial} counts to basic blocks and jumps in the function of $B_{new}$. In the example, three blocks
with hashes 0xA1, 0xB2, and 0xC3 are not modified; thus, we can compute and assign their counts.
Similarly, we assign initial counts for jumps whose both endpoints are matched based on the hashes.
The main challenge is to define the computation of the hash values. Very strict hashes that are
based on opcodes and operands of all block’s instructions are unstable to minor changes in the generated code,
such as adding nops or a different register allocation. Loose hashes are more stable but might result in collisions,
which is often hard to resolve correctly. Our strategy, described in detail in \cref{sect:matching}, is
to define a hierarchy of hashes ranging from the strictest (based on all relevant block's features) to
the weakest (based on a few features). This strategy allows matching unchanged blocks with high confidence using the strict hash
and, at the same time, provide some matches for blocks with modifications.
We note that a similar hash-based matching is utilized in the earlier work~\cite{WPM00}.
The difference is that we use the assigned counts as the initial guesses which can be modified at the second phase.

Once the initial counts are determined, we propagate the values through the graph and fill in the missing counts.
To this end, we extend a recent work of He, Mestre, Pupyrev, Wang, and Yu~\cite{HMPWY22} for profile inference. 
The goal of the subroutine is to guess
the control flow in a graph given a partial assignment of block and jump counts. The algorithm
propagates the counts along the control-flow graph to make it \df{consistent} and \df{realistic}.
The former objective ensures that the sum of incoming and outgoing counts for each non-terminal vertex is the same, as
in the actual program execution.
The latter objective is used to distribute the counts evenly when there are multiple equally likely solutions, such
as for the successors of block 0xC3 in \cref{fig:algo}. Our extension of the algorithm in \cite{HMPWY22} is based of computing the maximum flow of
minimum cost and described in \cref{sect:inference}.

\subsection{Matching}
\label{sect:matching}

The goal of the phase is to find matches between basic blocks within a pair of functions.
This is done by calculating a $64$-bit hash value for each block based on its code content. Recall that
basic blocks often change between two releases of a binary; hence, a naive one-to-one comparison of
the hashes will likely result in a very few matches. In order to accommodate minor changes and
match as many basic blocks as possible, we introduce multiple levels of hash computation.

\begin{itemize}[leftmargin=*]
	\item A \df{loose} hash of a basic block is based on all distinct instruction opcodes of
	the block; that is, instruction operands are excluded from the computation, as they may
	change from one version to another.
	The hash is computed by creating a string from lexicographically ordered instruction
	opcodes, which is then hashed with a machine-independent \texttt{xxHash}. We ignore
	pseudo and nop instructions as well as unconditional jumps, as they are often added or removed as a result
	of basic block reordering.

	\item A \df{strict} hash is based on all instruction opcodes and their operands.
	The computation is order-dependent, that is, two blocks have the same strict hash if and only if
	they are comprised of the same instructions and their order is the same. Similarly to the
	previous case, we ignore pseudo instructions and unconditional jumps in the computation.

	\item A \df{full} hash of a basic block is based on the block's strict hash combined with
	loose hashes of its successors and predecessors. Empirically, the value is useful to resolve
	collisions between blocks with identical code content that are often a result of
	the function inlining pass, which substitutes calls to the same function with (identical)
	basic blocks of the callee.
\end{itemize}

In addition to the three values defined above, we associate each basic block with the \df{offset} of
its address from the beginning of the function in the binary; the offset is guaranteed to be unique
across basic blocks of a function. Thus, we have four hash values for every block in the binary.
The values are lowered into $16$-bit integers and concatenated to form a $64$-bit hash value, which
is stored in the profile.

In order to find a match for a basic block from $P_{old}$ in the set of blocks corresponding to the
function in $B_{new}$, we first check if there is a matching block with the same full hash. At this
stage, the collisions (that is, blocks with the same hash value) are rare. If we find a matching
candidate, we stop the matching for the block. Otherwise, we try to find a match based on the strict
hash, followed by the loose hash. At these levels the probability of collisions are higher, and we
break ties by choosing the candidate whose offset is the closest to the offset of the considered basic block.
In addition to the rule above, we always match the first (entry) basic block from the profile
to the first block in the binary.
We emphasize that the hash definitions and the matching strategy is of a heuristic nature,
which has been tuned on several large-scale binaries described in \cref{sect:bench}.
It is likely that further fine-tuning is possible, e.g., by introducing more hash levels
and including extra code features, such as function calls. 
In this paper, we opted for simplicity of the computation rather than potentially more accurate but less
interpretable schemes based, for example, on machine learning.

\subsection{Inference}
\label{sect:inference}

The goal of the phase is to reconstruct the counts of all basic blocks and jumps,
while respecting, as much as possible, the initially assigned estimations. 
Denote $f(v) \ge 0$, $v \in V$ and $f(u, v) \ge 0, (u, v) \in E$
to be vertex and edge counts, respectively, that we seek to find. Recall that some but not necessarily all
vertices and edges have initial counts, denoted $\cnt(\cdot)$; the case when a block or a jump is missing
the initial count is indicated by $\cnt(\cdot) = \unknown$.

We want the counts to be consistent with a real execution of a function, which starts at the entry block,
then traverses the blocks in some order along the edges, and ends at an exit block. Thus, we introduce the
\df{flow conservation} rules:
\[
f(v) = \sum_{(u, v) \in E} f(u, v) = \sum_{(v, w) \in E} f(v, w)
\]
for all non-entry non-exit vertices $v \in V\setminus T^{*}, v \neq s^{*}$, and
\[
f(s^{*}) = \sum_{(s^{*}, u) \in E} f(s^{*}, u) \text{,\quad}
f(t^{*}) = \sum_{(u, t^{*}) \in E} f(u, t^{*})
\]
for the entry $s^{*}$ and for all exits $t^{*} \in T^{*}$, respectively.

We also want to preserve the initial estimates derived at the matching phase.
To measure how close the counts to the estimations, we
introduce the following objective:
\begin{equation}
	\label{eq:1}
\sum_{\substack{v \in V\\ \cnt(v) \neq \unknown}} cost\big(f(v), \cnt(v)\big) +
\sum_{\substack{(u, v) \in E\\ \cnt(u, v) \neq \unknown}} cost\big(f(u, v), \cnt(u, v)\big),
\end{equation}
where the sum is taken over all vertices and edges with an assigned count.
Here the term $cost(f, \cnt)$ penalizes the change of a count from $f$ to $\cnt$:
\[
cost(f, \cnt) =
\begin{cases}
	k_{inc} (f - \cnt) & \!\!\!\text{if } f \ge \cnt \\
	k_{dec} (\cnt - f) & \!\!\!\text{if } f < \cnt
\end{cases}
\]
Here $k_{inc}$ and $k_{dec}$ are non-negative penalty coefficients, whose exact values are determined
in Section~\ref{sect:count_quality}.

Overall, we seek to solve an optimization problem of finding consistent counts of vertices
and edges of a given control-flow graph, $G=(V, E)$, with prescribed count estimations,
$\cnt(v), v \in V$ and $\cnt(u, v), (u, v) \in E$, while minimizing objective~(\ref{eq:1}).
He et al.~\cite{HMPWY22} show how a variant of the problem, which has $\cnt(u, v) = \unknown$
for \emph{all} edges of $G$, can be mapped to an instance of the minimum-cost maximum flow problem, and
hence, can be solved optimally in polynomial time.
Observe that their algorithm can in addition
preserve (as a secondary objective) provided branch probabilities. We can use the feature to
evenly distribute the counts among the successors, when they do not have initial counts, such
as for the successors of block 0xC3 in \cref{fig:algo}.

Next we show how to reduce our problem to the
more restricted variant studied in \cite{HMPWY22}. To this end, we identify and subdivide all edges of $G$ that
have an initial count. That is, for all $(u, v) \in E$ with $\cnt(u, v) > 0$, we replace
the edge by a new vertex $w$ and two (directed) edges $(u, w)$ and $(w, v)$; then
we set $\cnt(w) = \cnt(u, v)$ and $\cnt(u, w) = \cnt(w, v) = \unknown$. For example, the
graph in \cref{fig:algo} (middle) is modified by subdividing two edges, (0xA1, 0xB2) and
(0xB2, 0xC3), and setting their initial counts to $300$ and $150$, respectively.
Note that the new graph contains
at most $|V|+|E|$ vertices and at most $2|E|$ edges, and only vertices may have initial counts.
It is easy to verify that the reduction preserves the optimality of solutions.
Therefore, we can apply the algorithm from~\cite{HMPWY22} to solve the inference problem.
We conclude that the work provides a near-linear time implementation of the
algorithm in practice; we verify the runtime in \cref{sect:runtime}.

\subsection{Implementation and Engineering}
\label{sect:impl}
Here we give an overview of profiles supported by BOLT.

\begin{figure}[!tb]
	\centering
	\includegraphics[page=2,width=0.48\columnwidth]{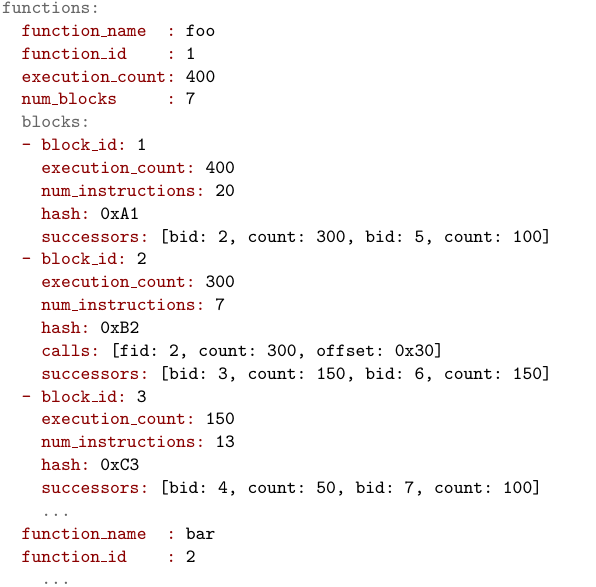}
	\caption{An example of the profile data in YAML format.}
	\label{fig:yaml}
\end{figure}

\begin{itemize}[leftmargin=*]
\item \textbf{Fdata profile} is an offset-based profile, which encodes taken control-flow 
	edges with their frequency and branch misprediction information. While
  the profile format is currently the default, it does not permit any 
  discrepancies between the binary it was
  collected from and the binary it is applied to. Therefore, it is unsuitable for
  the stale matching use case. 

	\item \textbf{Pre-aggregated perf profile} contains aggregated LBR data without
	binary knowledge. It encodes less information than fdata, as it lacks symbol
	information and omits fallthrough edges. It can be efficiently collected and stored
	but needs to be augmented with the binary it was collected on. This format
	is not self-contained and hence unsuitable for our use case.

\item \textbf{YAML profile} is a structured representation, which encodes control-flow 
	information along with function and basic block metadata. YAML profile explicitly 
	encodes basic blocks and
  control-flow edges and therefore, permits some discrepancies between the
  profiled and the input binaries. The format is the most flexible of all
  three and can easily be extended with necessary basic block metadata. 
  The primary downside is that processing YAML profile takes
  longer than fdata~---~in one representative case by about $70\%$.
  \cref{fig:yaml} illustrates the YAML format.
\end{itemize}

Stale matching is implemented as a part of the BOLT \texttt{YAML\-Profile\-Reader} class.
\textit{Matching} is abstracted from BOLT IR by basic block hashes, which
is added to YAML serialization format.
Basic block hashing is extended to capture necessary instruction and control-flow information.
\textit{Inference} is done on an IR-independent CFG representation (\texttt{FlowFunction}),
and reused in other parts of LLVM.
Stale profile matching is applied immediately after attaching the profile to CFG.
At this point, it is known if the function failed to match up with the profile.
Therefore, the algorithm works only with function profiles that would otherwise be discarded.

\section{Experimental Evaluation}
\label{sect:eval}

To validate the effectiveness of our approach, we (i) compare the performance of binaries generated with the use
of the new profile data, and (ii) evaluate the precision of the inferred block and jump frequencies.
The experiments presented in this section were conducted on a Linux-based server with
a dual-node 40-core 2.0 GHz Intel Xeon Gold 6138 (Skylake) having $256$GB RAM,
except \bin{chromium} which was conducted on a Linux-based desktop with
a 12-core 3.6 GHz Intel Core i7-12700K (Alder Lake) having $128$GB RAM.
The algorithms are implemented on top of \texttt{release\_16} of LLVM.

\subsection{Benchmarks}
\label{sect:bench}

We evaluated our approach on large open-source applications and real-world binaries
deployed at Meta's data centers; see Table~\ref{table:dataset}.
As publicly available benchmarks, we selected widely used programs that have large code size:
two open-source compilers (\bin{clang} and \bin{gcc}),
two database servers (\bin{rocksdb} and \bin{mysql}), and a browser (\bin{chromium}).
Next we discuss the testing setup for each of them.

\begin{itemize}[leftmargin=*]
	\item For \bin{clang}, we use the \texttt{release\_15} branch of LLVM as the base ($B_{new}$)
	release and the release with $3000$ commits behind that for $B_{old}$. They are built
	in \texttt{O3+\allowbreak LTO+\allowbreak AutoFDO} mode, and the profiles are collected by compiling several
	medium-sized template-heavy C++14 source files.

	\item For \bin{gcc}, we use \texttt{release-9.3.0} and \texttt{release-8.3.0} that were
	released approximately one year apart. We build the binaries with \texttt{O3+LTO} and collect
	the profiles on the same standalone benchmark of C++14 files.

	\item \bin{rocksdb} is a fast key-value storage; we utilize\linebreak \texttt{release-8.1.fb} and \texttt{release-7.1.fb}
	with ${\approx}1000$ commits from each other. The binaries are built
	in the \texttt{Release} mode with interprocedural optimization; \texttt{AutoFDO} is used to
	further speedup the binary.
	To collect profiles and measure performance, we run \texttt{db\_bench} on three built-in benchmarks,
	\texttt{fillrandom}, \texttt{fillseekseq}, and \texttt{overwrite}.

	\item Experiments with \bin{mysql} server are based on \texttt{mysql-\allowbreak 8.1.0} and \texttt{mysql-\allowbreak 8.0.28}
	released one year apart; we use the release version of the build using \texttt{O3+LTO} mode.
	We use \texttt{oltp\_read\_only} test from the \bin{sysbench} benchmark suite, running
	$4$ threads to read from the database initialized by inserting $100,000$ records into $8$ tables.

    \item For \bin{chromium}, we use milestones 111-115, that are aligned with
    Chrome releases. We utilize the \texttt{official} build configuration, which makes use
    of bundled \texttt{clang} and enables both \texttt{LTO} and \texttt{PGO}, with two tweaks:
    (i)~disabled control-flow integrity, and
    (ii)~disabled debug information generation to speed up builds.
    \texttt{Speedo\-meter\allowbreak2.0} is used to measure the performance of \bin{chro\-mi\-um}.
\end{itemize}

\begingroup
\setlength{\textfloatsep}{-2pt}
\setlength{\tabcolsep}{3pt}
\begin{table}[!tb]
	\footnotesize
	\centering
	\caption{Basic properties of evaluated binaries}
	\label{table:dataset}
	\begin{tabular}{@{\extracolsep{2pt}}lrrrrrrr@{}}
		\toprule
		\centering
		& \multicolumn{2}{p{5.8em}}{code size (MB)} & \multicolumn{2}{c}{functions} &
		\multicolumn{3}{c}{blocks per function} \\ \cline{2-3} \cline{4-5} \cline{6-8}
		& {hot} & {all} & {hot} & {all} & p50 & p95 & max \\
		\midrule
		\bin{clang}	  	  & $10.4$ & $83.2$ & $13,\!444$ & $95,\!943$  & $20$ & $274$ & $12,\!464$ \\
		\bin{gcc}	  	    & $3.6$  & $24.1$ & $6,\!971$  & $35,\!442$  & $28$ & $278$ & $5,\!974$  \\
		\bin{rocksdb}	    & $0.4$  & $5.1$  & $1,\!014$  & $9,\!610$   & $13$ & $238$ & $1,\!282$  \\
		\bin{mysql}	      & $1.5$  & $23.3$ & $2,\!501$  & $46,\!133$  & $6$  & $64$  & $5,\!152$  \\
		\bin{chromium}	  & $18.3$ & $171$  & $21,\!244$ & $541,\!722$ & $18$ & $151$ & $15,\!856$ \\
		\midrule
		\bin{hhvm}	  	  & $14$   & $230$ & $20,\!906$ & $476,\!861$     & $7$  & $62$  & $13,\!156$ \\
		\bin{dc1}	  	  & $28.5$ & $853$ & $56,\!578$ & $1,\!759,\!240$ & $13$ & $151$ & $9,\!773$  \\
		\bin{dc2}	  	  & $15.2$ & $892$ & $31,\!630$ & $2,\!146,\!062$ & $13$ & $156$ & $25,\!979$ \\
		\bin{dc3}	  	  & $17.7$ & $874$ & $45,\!256$ & $2,\!119,\!651$ & $13$ & $146$ & $20,\!379$ \\
		\bottomrule
	\end{tabular}
\end{table}
\endgroup

Note that while the intended use of our technique is with software deployments having short
release cycle, the open-source benchmarks have releases that are months apart. The aim of the
experiment is to show that the approach generalizes to the large number of accumulated differences.

For data-center workloads, we selected four binaries.
The first system is the HipHop Virtual Machine (\bin{hhvm}), that serves as an
execution engine for PHP at Meta, Wikipedia, and other large websites.
The other three binaries are large application running inside Meta's data centers. 
We use two consecutive releases for each of the binaries, which
are typically done on a weekly and bi-weekly basis. These binaries are built
with \bin{clang} in \texttt{O3+LTO} mode and use the compiler's PGO to enhance their performance.

\subsection{Binary Performance}
\begin{figure}[!tb]
	\centering
	\includegraphics[width=0.48\columnwidth]{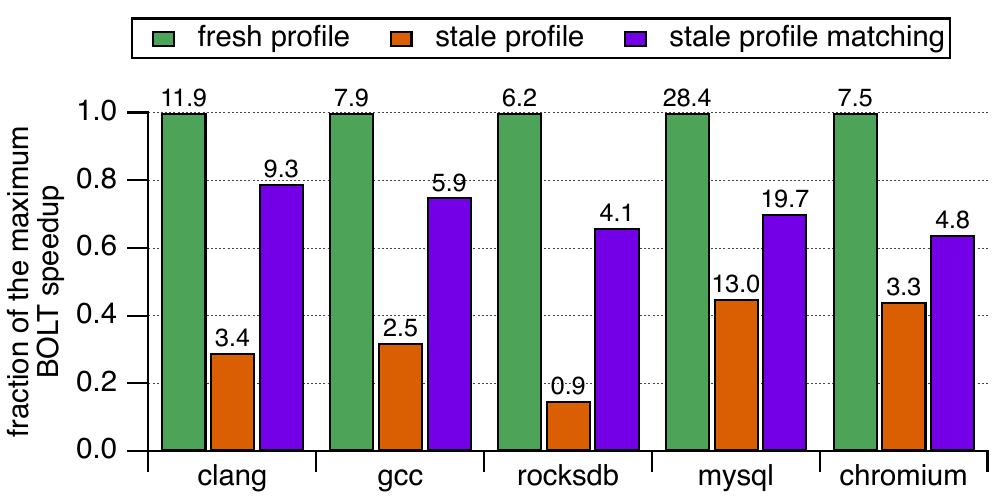}
	\caption{Relative and absolute performance improvements on the open-source benchmarks.
		The heights of the bars correspond to the fraction of the maximum performance improvement for each
		of the modes, while the values on top indicate absolute speedups.}
	\label{fig:perf_open}
\end{figure}

\cref{fig:perf_open} presents an evaluation on the open-source benchmarks. Here we compare
the following alternatives:

\begin{itemize}[leftmargin=*]
	\item the original non-BOLTed binary created by the compiler; this is our
	\alg{baseline} for the comparison;

	\item the existing strategy in BOLT that uses \alg{stale profile} discarding all
	functions with invalid profiles;
	
	\item an optimization using the most recent \alg{fresh profile} data; note that the 
	option is not practical and used only as an upper bound for optimizations;

	\item the newly proposed \alg{stale profile matching} approach.
\end{itemize}

For each benchmark, the figure shows the absolute speedup of the three strategies on top of \alg{baseline},
indicated by the number on top of the bars. The height of each bar is proportional to the fraction
of the maximum possible performance improvement achieved by each strategy. The values are obtained
by repeating each experiment $100$ times to increase the precision of the measurements so
that the average mean deviation is within $0.05\%$; thus, we omit the deviations from the figure.
We observe that \alg{stale profile matching} is able to recover $0.64{-}0.79$ of the maximum possible speedup
achieved by BOLT using fresh profile data. This is equivalent to $6.7\%$ (for \bin{mysql})
and $5.9\%$ (for \bin{clang}) absolute performance boosts on the benchmark.
In contrast, the existing strategy recovers only $0.15{-}0.45$ of the maximum speedup.

\begin{figure}[!tb]
	\centering
	\begin{subfigure}[t]{0.48\columnwidth}
		\includegraphics[width=0.98\columnwidth]{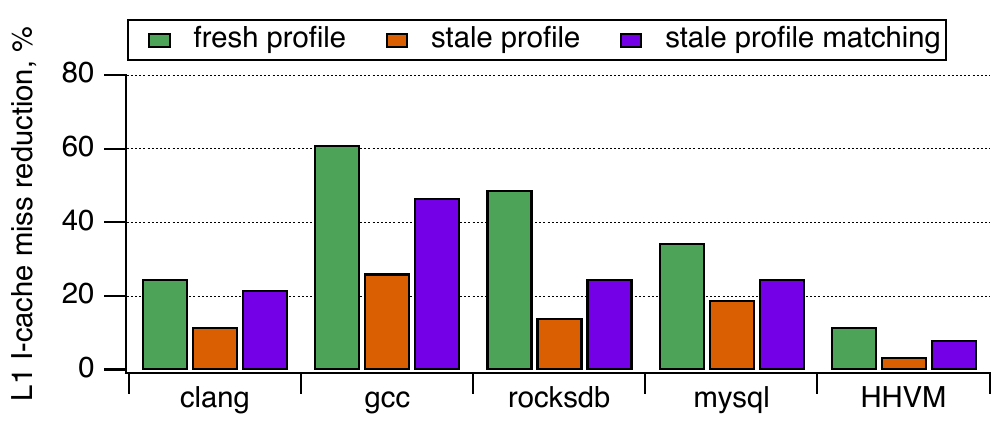}
	\end{subfigure}
	\hfill
	\begin{subfigure}[t]{0.48\columnwidth}
		\includegraphics[width=0.98\columnwidth]{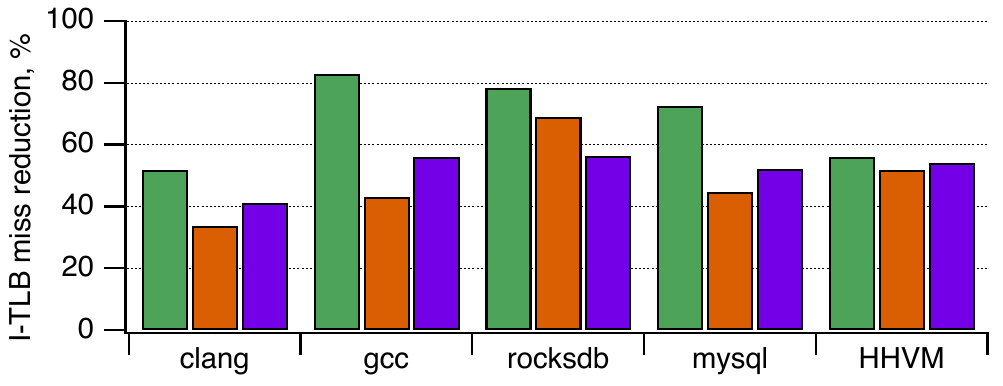}
	\end{subfigure}
	\caption{Improvements of the instruction cache \texttt{perf} metrics on top of non-BOLTed binaries.}
	\label{fig:metrics}
\end{figure}

To better understand the benefits of applying the new algorithm,
we collect several \texttt{perf} metrics related to code layout. As expected, the main
advantage of the algorithm is an improved performance of the L1 instruction and I-TLB caches;
refer to \cref{fig:metrics} where we use \texttt{front\-end\_re\-ti\-red.\allowbreak l1i\_miss}
and \texttt{front\-end\_re\-ti\-red.\allowbreak itlb\_miss} hardware events for estimating I-cache and I-TLB misses, respectively.
We also see a modest improvement in the number of branch misses and the
performance of the last level cache, though the absolute difference is less prominent.

In order to measure the impact of stale profile matching on production workloads, 
we use internal performance measurement tools for running A/B experiments at Meta.
We stress that the obtained measurements on the
workloads are noticeably noisier than on the standalone open-source benchmarks; empirically
typical deviations are up to $0.4\%$ for \bin{dc} binaries and up to $0.2\%$ for \bin{hhvm}.
For \bin{hhvm}, we record a $5.7\% {\pm} 0.2\%$ performance improvement by using \alg{fresh profile},
which degrades to a $3.1\%$ value for \alg{stale profile} containing $64\%$ of stale samples.
The new \alg{stale profile matching} achieves $4.5\%$ performance boost on top of \alg{baseline}, 
which is equivalent to recovering $0.79$ of the maximum speedup.
For \bin{dc} services, the existing experimentation infrastructure only allows to measure
absolute speedups of the new approach over \alg{stale profile}. We record a
$0.2\% {\pm} 0.4\%$ boost for \bin{dc1} containing $14\%$ stale samples, 
$0.5\% {\pm} 0.3\%$ boost for \bin{dc2} containing $21\%$ stale samples, and
$0.6\% {\pm} 0.3\%$ boost for \bin{dc3} containing $29\%$ stale samples.
That is, for three out of four production workloads, we are able to measure a statistically significant 
improvement by turning on the technique.

Finally, we discuss the performance of \alg{stale profile matching} with respect to
alternative solutions. A recent work of Moreira et al.~\cite{PMO23} provides
an evaluation of their technique, referred to as Beetle, as well as the earlier
approach, BMAT, by Wang et al.~\cite{WPM00}. In the evaluation, Moreira et al.~\cite{PMO23}
estimates the impact of Beetle and BMAT on a number of open-source binaries. We observe that
their evaluation differs from our setup described in \cref{sect:bench} in that it does not
employ a compiler's PGO, and hence, a direct comparison of the results is impossible. To mimic
the setup of \cite{PMO23}, we repeat the experiment with the \bin{clang} binary by building
its \texttt{release\_14} with \texttt{O3+LTO}, while using \texttt{release\_10} for collecting
profile data. Similar to~\cite{PMO23}, we record $75\%$ of stale functions in the profile, and
the speedups achieved by \alg{fresh profile} and \alg{stale profile} are $28\%$ and $6\%$, respectively; 
all three measurements closely match the reported values in the paper.
Moreira et al.~\cite{PMO23} provide evaluations of multiple techniques on the benchmark, 
including Beetle and BMAT, and none of them exceeds a $7\%$ speedup on top of non-BOLTed binary. In contrast, 
\alg{stale profile matching} yields a substantially higher $18\%$ performance boost.
Analogously, \cite{PMO23} reports the mean recovered fraction of the maximum BOLT speedup
on several benchmarks to be $0.25$ for BMAT and $0.43$ for Beetle,
while \alg{stale profile matching} recovers $0.6{-}0.8$ of the speedup on the arguably more
challenging \texttt{O3+LTO+AutoFDO} setup.
Therefore, we do not consider earlier studies in \cite{WPM00,PMO23} as viable alternatives
for the problem.

\subsection{Quality of Inferred Profiles}
\label{sect:count_quality}

To evaluate the accuracy of the algorithm, we compare the resulting profile data of
with \alg{fresh profile}. To this end, we employ two
measures that quantify the similarity between the two profiles. The first one
follows earlier works on profile inference and referred to as the \df{edge overlap}~\cite{Chen13,HMPWY22}.
Let $f(e), e \in E$ be the constructed edge counts by the algorithm, and
$gt(e), e \in E$ be the counts of the edges in the fresh profile.
The \df{edge overlap} is defined as
\[
\sum_{e \in E} \min\Big(\frac{f(e)}{\sum_{e \in E} f(e)}, \frac{gt(e)}{\sum_{e \in E} gt(e)}\Big),
\]
where the sum is taken over all edges of the graph. If the counts in the two profiles match exactly, the overlap is 
equal to $1$; otherwise, the measure takes values between $0$ and $1$.
The second measure is called the \df{tsp score}. It is motivated by studies on code layout~\cite{PH90,MPU21,NP20}
in which co-locating basic blocks that frequently call each other is beneficial for the performance.
To compute the value, we apply basic block reordering for all functions in a binary based
on the fresh profile, and calculate the sum of counts for all \df{fallthrough} edges, that is, 
edges between pairs of consecutive blocks; denote the value by $\tsp(gt)$. Next, we compute 
$\tsp(f)$, where we use inferred counts, $f$, for basic block reordering but fresh counts for the summation.
Observe that $\tsp(gt) \ge \tsp(f)$ for the
majority of instances since existing algorithms for basic block reordering produce near-optimal
layouts with respect to the measure~\cite{NP20,MPU21}. The \df{tsp score} is defined by
$0 \le \frac{\tsp(f)}{\tsp(gt)} \le 1$ and estimates how suitable are the constructed profiles
for code layout.

\begin{figure}[!tb]
	\centering
	\includegraphics[width=0.48\columnwidth]{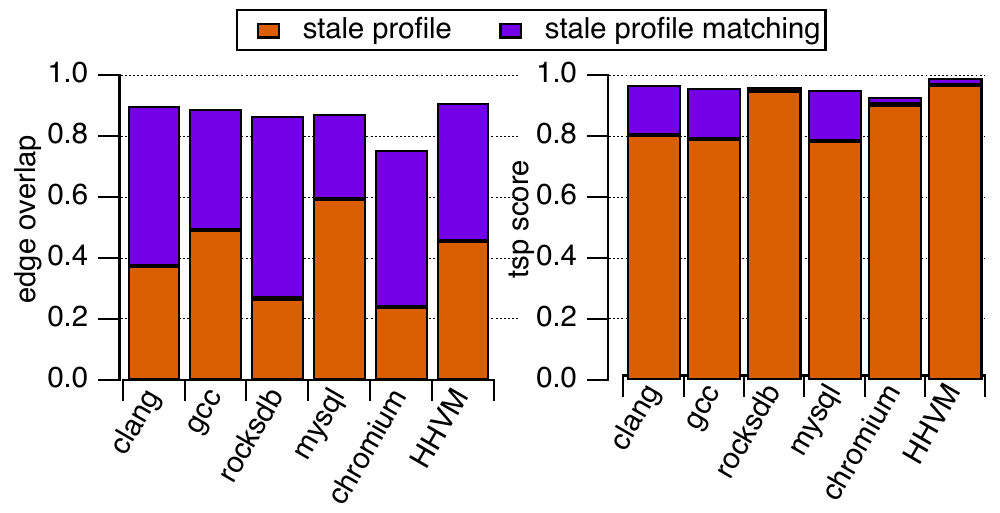}
	\caption{Edge overlap and tsp score for the benchmarks.}
	\label{fig:proxy}
\end{figure}

\cref{fig:proxy} demonstrates the analysis of the two measures on open-source benchmarks and \bin{hhvm}.
Edge overlap values range from $0.76$ (for \bin{chromium}) to $0.91$ (for \bin{clang}, \bin{gcc}, and \bin{hhvm}), which
is a substantial improvement over $0.24{-}0.6$ for the stale data. 
For the \df{tsp score}, we record an improvement from $0.8{-}0.9$ to $0.93{-}0.96$ on
the open-source benchmarks and to the value of $0.99$ on \bin{hhvm}. Loosely speaking, the
latter result indicates that in $99\%$ of instances, the generated profile is as good for
basic block reordering as the fresh one.
We summarize the findings by noting that \alg{stale profile matching} is able to significantly
reduce but not completely eliminate the gap to the fresh profile data.

The primary parameters of \alg{stale profile matching} are penalty coefficients, 
$k_{inc}$ and $k_{dec}$, introduced in Section~\ref{sect:inference}.
Using the \df{edge overlap} score, we identified the following combination resulting in the
highest value: $k_{inc} = 1$ and $k_{dec} = 2$. Intuitively, it is twice more expensive to 
decrease a block count from its initial estimate than to increase it.

\subsection{Build Time}
\label{sect:runtime}
\begin{figure}[!tb]
	\centering
	\includegraphics[width=0.45\columnwidth]{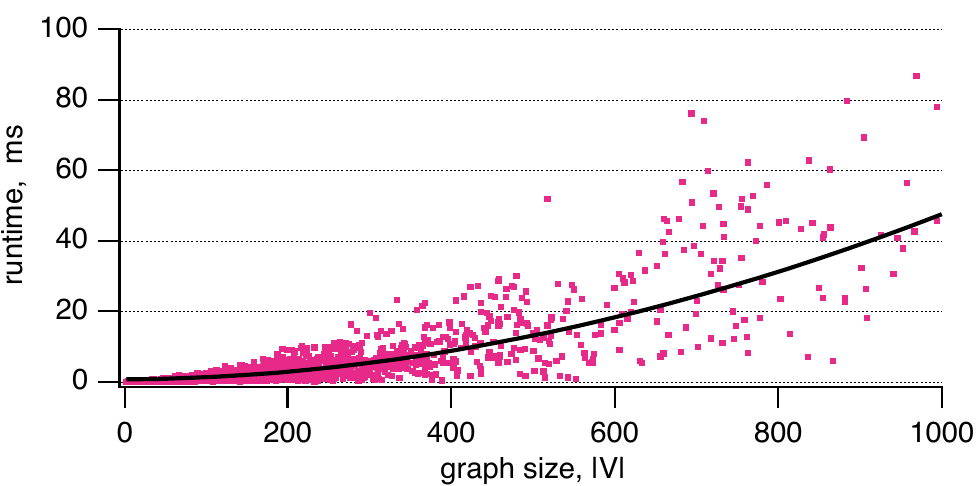}
	\caption{The runtime (in milliseconds) of \alg{profi} as a function of the number of basic 
		blocks measured on \bin{hhvm}.}
	\label{fig:runtime}
\end{figure}

Here we evaluate the runtime of \alg{stale profile matching} and its impact on the overall BOLT's
processing time. The absolute running times (in milliseconds) of the algorithm are illustrated in \cref{fig:runtime}. While the
observed complexity of the algorithm is super-linear in the number of basic blocks, the
runtime does not exceed $0.1$ second even for instances containing $|V|=1000$ blocks. 
The majority of real-world instances contain much fewer vertices (see \cref{table:dataset}), and the
largest recorded runtime across all functions in our dataset is $0.8$ second. Hence, our algorithm
is unlikely to introduce an overhead to the existing build process of data-center applications.
The total time taken by \alg{stale profile matching} for $21$K functions of \bin{hhvm}
is under $20$ seconds, which is only a small fraction of the overall processing time
for the binary~\cite{PASO21}.

\section{Related Work}
\label{sect:related}
There exists a rich literature on profile-guided compiler optimizations.
Traditionally profile data is used for function inlining and outlining~\cite{LRN22,DPGPR21},
basic block reordering~\cite{PH90,NP20,LCD19}, function layout~\cite{PH90,OM17,HLMP23},
merging similar functions~\cite{RPWCL20,RPWC21},
loop optimization~\cite{RPFB22}, register allocation~\cite{LCBS19,FSW23}, and many others.
These techniques are implemented in a variety of compilers and binary optimizers
used for static and dynamic languages~\cite{Ott18,OL21}, and
can be applied at the compile time~\cite{CML16}, link time~\cite{SDAL01,LFZLS22}, 
or post-link time~\cite{LMPCL04,PANO19,Prop21}.

Most of the above tools assume that a representative profile dataset is available. In practice, however,
it is often difficult to generate an adequate profile even for a binary processing a specific benchmark.
Chen et al.~\cite{Chen13} observe hardware-related problems that lead to inaccurate profiles
and suggest several tricks to mitigate the hardware bias. Later studies explore alternative
profile improvement techniques, such as varying the sampling rate~\cite{WZSGSY13} and utilizing
LBR technology available on modern Intel x86 processors~\cite{NYMZ15,CML16}.
However, not all processor vendors provide such logging functionality, and moreover, recent
studies indicate that LBRs still suffer from sampling bias~\cite{XWSJL19,YDDC20}.
Therefore, it is common to apply a post-processing step to correct the profiles\cite{LNH08,HMPWY22}. 
An alternative direction to improve profile accuracy is to employ machine learning
to predict the frequencies of the blocks, functions, and branch probabilities~\cite{BL93,WL94}.
Despite a wealth of recent studies in the area~\cite{RC21,MOQ21,RL22}, we are not
aware of a successful application of such a technique at scale in production environment.

The problem explored in this paper is complementary to the above studies and has been mostly
overlooked by the community. In order to reuse profile data collected on previous releases of
an application, the BMAT system by Wang et al.~\cite{WPM00} uses a hash-based
matching between basic blocks. Their approach is equivalent to the first phase of our
algorithm; this phase alone, as we show in the experimental evaluation, is not competitive with
the two-phase approach. Another closely related (and still unpublished) work is the
Beetle technique by Moreira et al.~\cite{PMO23}. The authors suggest to extract
certain branch characteristics (e.g., direction and opcode) and use them to map
profile information across releases. Similarly to our study, the work~\cite{PMO23} is
evaluated in BOLT. \cref{sect:eval} indicates that our approach achieves
substantial performance gains over Beetle.
Finally, we mention another recent stream of work that might mitigate the profile staleness:
an online (or just-in-time) system for optimizing binaries written in unmanaged languages.
In that scenario, the binary starts running in an unoptimized state and its profile is collected during a
certain warm-up period. Once enough profile data is collected, we apply optimizations and shift the execution
to the new code. Ocolos is one recent prototype implementing such an approach~\cite{ZKPKLD22}.
While the profile staleness problem is eliminated, this approach introduces significant profiling and optimization
overhead that could be amortized for long-running binaries.

More generally, the problem of inferring stale profiles can be seen as a special variant
of binary code similarity, whose goal is to identify differences and similarities between two
pieces of binary code. This is fundamental task for many applications;
refer to the recent surveys on the topic~\cite{Alr20,HC21}.

\section{Conclusion}
\label{sect:conclusion}

We designed and implemented a novel approach for re-using profile data
collected on binaries built from several revisions behind the release. Based on an extensive
evaluation, we conclude that the proposed technique can recover a large portion of the performance loss due to
outdated profiles. One direct implication is a simplified maintenance of
performance-critical applications that are frequently released; by introducing stale
profile matching, such applications can be released without re-collecting profiles
after every hotfix.
A possible future direction is to adapt our implementation in BOLT to other PGO tools, 
such as Propeller~\cite{Prop21} or AutoFDO~\cite{CML16}. While we do not foresee any
high-level obstacles, there might be non-trivial implementation challenges.

From the theoretical point of view, our problem can be considered as a variant of
graph alignment or graph matching problems~\cite{HC21}. Many of the corresponding 
problems are computationally hard for general instances. However, real-world control-flow
graphs arising in applications exhibit certain structural properties, which may lead to
algorithms with provable guarantees.

\balance

\bibliographystyle{ACM-Reference-Format}
\bibliography{stale_refs}

\end{document}